\documentclass[preprint,12pt]{elsarticle}

\usepackage{amssymb}
\usepackage{color,soul}

\newcounter{bla}

\usepackage{algorithmic}
\usepackage{varwidth}
\usepackage{multirow}
\usepackage{amsmath}
\usepackage{listings}
\journal{Computers \& Fluids}

\begin{document}

\begin{frontmatter}



\title{Energy efficiency of finite difference algorithms on multicore CPUs, GPUs, and Intel Xeon Phi processors}


\author[a]{Satya P. Jammy\corref{cor1}}
\author[a]{Christian T. Jacobs}
\author[a]{David J. Lusher}
\author[a]{Neil D. Sandham}

\cortext[cor1] {Corresponding author.\\\textit{E-mail address:} s.p.jammy@soton.ac.uk}
\address[a]{Aerodynamics and Flight Mechanics Group, Faculty of Engineering and the Environment, University of Southampton, University Road, Southampton, SO17 1BJ, United Kingdom}

\begin{abstract}
In addition to hardware wall-time restrictions commonly seen in high-performance computing systems,
it is likely that future systems will also be constrained by energy budgets. In the present work, 
finite difference algorithms of varying
computational and memory intensity are evaluated with respect to both energy efficiency and
runtime on an Intel Ivy Bridge CPU node, an Intel Xeon Phi Knights Landing processor,
and an NVIDIA Tesla K40c GPU. The conventional
way of storing the discretised derivatives to global arrays for solution advancement is found to be inefficient in terms
of energy consumption and runtime. In contrast, a class of algorithms in which
the discretised derivatives are evaluated on-the-fly or stored as thread-/process-local variables
(yielding high compute intensity)
is optimal both with respect to energy consumption and runtime.
On all three hardware architectures considered, a speed-up of $\sim 2$ and an energy saving of $\sim 2$ are observed for the high
compute intensive algorithms compared to the memory intensive algorithm. The energy consumption is found to be proportional to runtime, irrespective of the power consumed and the GPU has an energy saving of $\sim$ 5 compared to the same algorithm on a CPU node.
\end{abstract}


\end{frontmatter}

% \linenumbers

\section{Introduction}\label{sect:introduction}
As the performance of high-performance computing (HPC) systems tends towards exascale, the characteristics
of hardware are not shaped by floating-point operation (FLOP) count and memory bandwidth alone. Energy costs continue to rise in an
era where climate change is a factor in many decision making processes, and power consumption and energy
efficiency are therefore prominent concerns of hardware manufacturers and end-user software developers
alike. According to a report by the US Department of Energy \citep{Ashby_etal_2010}, power consumption
of a potential exascale machine is predicted to increase by a factor of three relative to HPC system
designs in 2010. Indeed, an exascale-capable machine built with the technology existing in 2010 would
be expected to cost more than \$2.5 billion USD in power expenditure alone \citep{Ashby_etal_2010}
and consume at least a gigawatt of power \citep{Miller_2010, Slotnick_etal_2014}.

The need for improved energy efficiency in HPC has led to the Green500 list \cite{Feng2007}. 
In addition to listing the raw
compute capability of a system, Green500 ranks systems by their power efficiency via the FLOPS/Watt
metric. The impact of heterogeneous systems which combine CPU and accelerators is clear to see. 
Despite the improvements brought about by new architectures, a 2013 review of the Green500 list
\citep{Subramaniam2013, Mair2015} concluded that the goal of a 20MW exaflop machine in 2020 by
the Defence Advanced Research Projects Agency (DARPA) would not be met with current hardware trends.
A machine of this standard would require 1000 times the performance of a current petascale system,
but at equivalent levels of power consumption.

Graphical Processing Unit (GPU) and Many Integrated Core (MIC)-based architectures, have augmented conventional
CPU-based systems and are currently a popular choice to increase the FLOP count.
However there is uncertainty in the future of architecture design for HPC machines.

From the perspective of numerical simulations, it is quite probable that in addition to the hardware
wall-time restrictions commonly seen in institutional, national and international HPC systems,
simulations will also be constrained by energy budgets. It is therefore paramount that numerical codes and their underlying solution algorithms can be optimised in both respects. However,
while such improvements can be made on both a hardware and a software level, these can often entail extensive
re-writes of the codebase. Drastic changes may be required in order to target a different backend Application Programming
Interface (API) and introduce optimisations for a particular piece of hardware.

In the approach to exascale computing, recent research has looked at the energy efficiency
of various codes and architectures (see e.g. \cite{McIntosh-Smith_2012}).
The energy efficiency of sparse matrix multiplication on a GPU,
Intel Xeon Phi and Field Programmable Gate Arrays (FPGAs) was investigated by \cite{Giefers2016}. The FPGAs and accelerators were found to have excellent energy
efficiency for computing sparse matrix kernels, suggesting that off-loading certain tasks to specialised hardware
components could be a viable option in future systems.

Matrix multiplication benchmarks offer some insight into the energy efficiency of different architectures,
but are not always representative of large-scale codes comprised of multiple components and different workloads.
Energy efficiency studies of large-scale scientific codes for N-body molecular dynamics simulations have been investigated
by \citep{Zecena2013, Brown2017}. They concluded that any reduction in runtime gives proportional energy savings.

The use of low-power Advanced RISC Machine (ARM)-based processors in the context of fluid dynamics simulations with spectral element,
finite element and Lattice-Boltzmann-based codes was investigated by \cite{Goeddeke_etal_2013}, focussing on the trade-off between time-to-solution
and energy-to-solution. They
found that the ARM processors are more energy efficient than traditional x86-based CPUs.

The energy saving of a second-order 3D finite-difference code for solving the
acoustic diffusion equation (a homogeneous parabolic partial differential equation) on an Intel Xeon CPU, Intel Xeon Phi
co-processor and two NVIDIA GPU cards compared to a sequential implementation has been investigated by \cite{Hernandez2015}.
They concluded that the energy consumed
by the GPU is the lowest and CPU is highest, with the Phi in between. They also concluded that
energy consumption depends on the runtime of the simulations. Techniques to further improve energy efficiency
for GPU architectures (e.g. better use of caches and improved CPU--GPU workload division) are surveyed in \cite{Ashby_etal_2010}.

Besides changing the architectures it is also possible to change the programming model. A comparison of parallel programming paradigms (OpenCL, OpenACC, OpenMP and CUDA) was studied by
\cite{Memeti:2017:BOO:3110355.3110356}, assessed on their performance and energy consumption on a
variety of architectures. For each application considered, the execution time was the
dominant factor in the overall energy consumption, with implementations offering the fastest runtime
once again yielding greater energy efficiency.

Finite difference methods for the solution of partial differential equations are used in many areas of
science and engineering, such as wave propagation modelling, acoustic diffusion, heat conduction,
fluid dynamics, micro magnetics, seismic imaging, plasma flows, electromagnetics, and
magnetohydrodynamics (MHD) flows to name a few.
Typically large multi-dimensional systems of partial differential equations (PDE) need to be solved.
Recent effort in optimising such stencil-based codes has focussed on designing efficient
solvers with the help of tiling, and improving data locality, etc. For example, the reader is referred to
\cite{Tang_etal_2011, Christen_etal_2011} and the references therein. Most of these works involve optimising
the existing codebases, however it is also possible to introduce more radical algorithmic changes.
Recent work \citep{Jammy_etal_InPress} using the automatic source code generation capabilities of the
OpenSBLI framework \citep{Jacobs_etal_2017} has demonstrated that
varying the computational and memory intensity of explicit finite difference algorithms for solving the compressible
Navier-Stokes equations in three dimensions on a representative test case, has a significant impact
on simulation runtime on an Intel Ivy Bridge CPU node. However, not much is known about how these algorithms behave across various architectures with respect to
their energy efficiency and power consumption.

This paper investigates the performance, in terms of both runtime and energy
efficiency, of six algorithms (five of which are from \citep{Jammy_etal_InPress}) on an Intel Ivy Bridge CPU node, an NVIDIA K40c GPU, and an Intel Xeon Phi Knights Landing processor (denoted KNL for brevity from here on in) for a compressible Navier-Stokes test case. Section \ref{sect:methodology} introduces the numerical method and the algorithms
under consideration, along with the automatic source code generation framework (OpenSBLI)
used to generate C code implementations for each algorithm. In Section \ref{sect:hardware}, a description of the various backend programming models and the hardware used is given, along with a brief discussion of how the power consumption and energy usage were measured. The results are presented in Section \ref{sect:results} and demonstrate that the algorithms
requiring minimal memory access are consistently the best-performing algorithms across the three
different architectures. The paper closes with some concluding remarks in Section \ref{sect:conclusion}.

\section{Methodology}\label{sect:methodology}

The governing equations are the non-dimensional compressible Navier-Stokes equations with constant viscosity, given by

\begin{equation}
\frac{\partial \rho}{\partial t} =  -\frac{1}{2} \left( \frac{\partial \rho u_j }{\partial x_j}
+ \rho \frac{\partial u_j }{\partial x_j} + u_{j} \frac{\partial \rho }{\partial x_j} \right),
\label{mass}
\end{equation}
\begin{equation}
\frac{\partial \rho u_i }{\partial t}  = -\frac{1}{2} \left( \frac{\partial \rho u_i u_j }{\partial x_j}
+ \rho u_i \frac{\partial u_j }{\partial x_j} + u_{j} \frac{\partial \rho  u_i}{\partial x_j} \right)
- \frac{\partial p}{\partial x_i}  +\\
\frac{1}{\mathrm{Re}}  \left(\frac{\partial^{2} u_i}{\partial x_j^{2}}   + \frac{\partial^{2} u_j}{\partial x_i\partial x_j}   - \frac{2}{3}  \frac{\partial^{2}  u_k}{\partial x_j\partial x_k}
\delta_{i j}\right),
\end{equation}
and
\begin{eqnarray}
\frac{\partial \rho E}{\partial t}   = - \frac{1}{2}  \left(
    \rho E {\frac{\partial u_j}{\partial x_j}}  - u_j{\frac{\partial \rho E }{\partial x_j}
}  - \frac{\partial \left(\rho E u_j\right)}{\partial x_j} \right)  - \frac{\partial \left(p u_j\right)}{\partial x_j} + \\ \nonumber
\frac{u_i}{\mathrm{Re}}  \left(\frac{\partial^{2} u_i}{\partial x_j^{2}}  +
\frac{\partial^{2} u_j}{\partial x_i\partial x_j} - \frac{2}{3}
\frac{\partial^{2} u_k}{\partial x_j\partial x_k}  \delta_{i j}\right) + \\ \nonumber
\frac{1}{\mathrm{Re}} {\frac{\partial u_i}{\partial x_j}}  \left(\frac{\partial u_i}{\partial x_j}  + \frac{\partial u_j}{\partial x_i}  -
\frac{2}{3}  \frac{\partial}{\partial x_k} u_k  \delta_{i j}\right) +\\ \nonumber
\frac{1}{\left(\gamma - 1\right) \mathrm{M} ^{2} \mathrm{Pr} \mathrm{Re} } {\frac{\partial^{2} T}{\partial x_j^{2}}},
\label{energy}
\end{eqnarray}

\noindent for the conservation of mass, momentum and energy, respectively. Note that the convective terms are written in the skew-symmetric formulation of \citep{Blaisdell1996} and the Laplacian form of the viscous terms is used to improve the stability of the equations \citep{Pirozzoli2015, Salvadore2013, Sandham2002}. 
Subscripts are the Einstein indices and repeated indices imply summation.
The quantity $\rho$ is the fluid density, $u_i$ is the velocity vector, $p$ is pressure, $E$ the total energy,  $\mathrm{Re}$ is the Reynolds number, $T$ is temperature, $\mathrm{M}$ is the Mach number, $\mathrm{Pr}$ is the Prandtl number, $\gamma$ is the ratio of specific heats and $x_i$ is the coordinate system.

The pressure and temperature relations are given by,
\begin{equation}
\label{eq:pinprimitive}
p = \left(\gamma -1 \right)\left(\rho E - \frac{1}{2} \rho u_j^{2}\right),
\end{equation}
and
\begin{equation}
T = \frac{\gamma \mathrm{M}^{2} p}{\rho},
\label{eq:Tinprimitive}
\end{equation}
respectively.

\subsection{Spatial discretisation}
The spatial derivatives in the governing equations are evaluated using a fourth-order central
finite-difference method. The first and second order partial derivatives of an arbitrary function
($f$) in the $x_0$ coordinate direction, with a grid spacing of {$h_0$} are given by,
\begin{equation}
\frac{\partial f} {\partial x_0} = \frac{1}{h_0}\left(\frac{f_{i-2}}{12} - \frac{2f_{i-1}}{3} + \frac{2f_{i+1}}{3} - \frac{f_{i+2}}{12} \right),
\label{eq:fd}
\end{equation}

\begin{equation}
\frac{\partial^{2} f} {\partial x_0^{2}} = \frac{1}{h_0^{2}}\left(-\frac{f_{i-2}}{12} + \frac{4f_{i-1}}{3} - \frac{5f_{i}}{2} + \frac{4f_{i+1}}{3} - \frac{f_{i+2}}{12} \right),
\label{eq:sd}
\end{equation}
where $i$ is now the grid point's index. The numerical derivatives of the function in the other coordinate directions can be evaluated in a similar fashion. As periodic boundary conditions are used in the present work, no special treatment of the grid points on the boundary are required.

\subsection{Temporal discretisation}
The variables that are advanced in time ($\vec Q = [\rho, \rho u_i, \rho E] ^{T}$) are referred to as the solution variables.
In the present work a three-stage low storage Runge-Kutta method is used to advance the solution
variables in time; this can be written as

\begin{align}\label{rk3_tvd}
\vec{Q}^{1} &= \vec{Q}^{n},\\ \nonumber
\vec{Q}^{k} &= \vec{Q}^{k-1} + \Delta t R\left(\vec{Q}^{k}\right),
\end{align}
where $R(\vec Q)$ represents the numerical evaluation of the spatial derivatives (referred to as the residual), $n$ is the time-level, $\Delta t$ is the timestep, $k$ = 1,2,3 is the sub-stage of the Runge-Kutta timestepping loop, and the solution vector at the next time-level ($n$+1) is $\vec Q^{n+1} = \vec{Q}^{3}$.

\subsection{Algorithms}
Pseudo-code for the spatial and temporal discretisation of the governing
equations is outlined in Figure \ref{fig:pseudocode}. Most of the computational time is spent evaluating the residual. This consists of
(a) evaluating the primitive variables ($u_i, P, T$), (b) evaluating the spatial derivatives using equations \ref{eq:fd} and \ref{eq:sd} (63 derivatives are to be computed), and (c) to compute the residual of the equations.
The algorithms considered in this work differ in the way the spatial derivatives are evaluated and the
residuals are computed. The other steps are evaluated the same way across all the algorithms, as in \citep{Jammy_etal_InPress}.
Note that all the steps in Figure \ref{fig:pseudocode} are applied on the entire domain (grid size) except
the application of periodic boundary conditions, which are applied over each boundary face.

\begin{figure}[h]
\noindent\fbox{%
\begin{varwidth}{\dimexpr\linewidth-2\fboxsep-2\fboxrule\relax}
\begin{algorithmic}
\STATE \texttt{initialise-the-solution-variables}
\FOR{\texttt{each-timestep}} \STATE {\texttt{save-state ($\vec Q^{n}$)} }
\FOR{\texttt{each-runge-kutta-substep}}
\STATE {\texttt{evaluate $R(\vec Q^{k})$}}
\STATE {\texttt{evaluate $\vec Q^{k+1}$}}
\STATE {\texttt{n+1-level-solution-advancement $\vec Q^{n+1}$}}
\STATE {\texttt{apply-periodic-boundary}}
\ENDFOR \texttt{  //  runge-kutta-substep}
\ENDFOR \texttt{  //  timestep}
\end{algorithmic}
\end{varwidth}
}
\caption{Generic algorithm for the numerical simulation of the compressible Navier-Stokes equations
using a three stage Runge-Kutta timestepping scheme.}
\label{fig:pseudocode}
\end{figure}

The different algorithms are outlined below. The source code for the implementation of each algorithm is generated automatically using the OpenSBLI framework \cite{Jacobs_etal_2017}.

\subsubsection{Baseline (BL)}
The baseline algorithm follows the conventional approach frequently used in large-scale finite
difference codes \cite{Sandham2002}.  For example, to evaluate the derivative of a 3D array (\texttt{f})
in the $x_0$ coordinate direction, a typical Fortran implementation would look like:

\begin{lstlisting}[caption = {Fortran 90 subroutine to evaluate the partial derivative of a function
in the x0 coordinate direction.}, ,captionpos=b]

subroutine derx(f, df)
  real :: factx
  integer :: i,j,k
  factx = 1.0/(12*h0)
  do k=1,nx2
    do j=1,nx1
      do i=1, nx0
        df(i,j,k) = factx*(8.0*(f(i+1,j,k) - f(i-1,j,k))
                    - f(i+2,j,k) + f(i-2,j,k))
      end do
    end do
  end do
endsubroutine derx
\end{lstlisting}
% caption={}
This subroutine takes any arbitrary array \texttt{f} as input and the derivative is evaluated by
looping over all the grid points (nx0, nx1, nx2) in the domain by applying equation \ref{eq:fd} and
stored to the global `work array' \texttt{df}.

Due to ease of programming such individual subroutines are typically written for the first and second derivatives in different coordinate directions.
In order to evaluate the 63 derivatives in the
equations, a subroutine is called depending on the coordinate direction and order of the derivative,
and the output is stored into arrays. Some of the derivatives in the equations require a combination of variables,
(for example, $\partial (\rho u_2 * u_0)/ \partial x_0$); first the variable $\rho u_2 * u_0$ is evaluated
into a temporary array, which is then passed as an input to the subroutine to obtain the derivative and store it to an array. Once all the spatial  derivatives are evaluated and stored into arrays, the residual of the
equation is computed by replacing the continuous derivatives with their corresponding arrays.

The implementation of the BL algorithm in OpenSBLI follows this approach, and requires (without any reuse of arrays)
$63$ arrays for the derivatives. This is the least computationally intensive algorithm that is tested.

A variant of this algorithm was considered in order to reduce the number of arrays used. This was done
by splitting up the spatial terms in the governing equations into two groups: The terms that feature the Reynolds number Re as one group and
the remaining terms as another. The residuals are evaluated
by adding the residual of each individual group. This version
resulted in a reduction in the number of arrays to 40. However, as the reduction in runtime was less than 2\% and
the optimisations being specific to the compressible Navier-Stokes equations, they are not considered in the present work.

\subsection{Intermediate storage algorithms}
Two variants of the BL algorithm that reduce the number of arrays by increasing the computational intensity
of the algorithm were considered. As the components of the velocity gradient tensor ($\partial u_i/\partial x_j$) are reused across different equations,
these 9 derivatives are evaluated in a similar fashion to the BL algorithm (i.e. calling
separate subroutines and storing the derivative to an array).
For the remaining spatial derivatives, two different options are considered, resulting in the following algorithms.

\subsubsection{Recompute Some (RS)}
In the RS algorithm the remaining spatial derivatives are evaluated on-the-fly directly in the residual. For example, consider
the residual (right hand side) for equation \ref{mass}. In the RS algorithm the
remaining derivatives are replaced by their finite-difference stencil (equation \ref{eq:fd}) which is evaluated on-the-fly, as opposed to evaluating the derivative to an array for later retrieval. In other words, the derivatives that are not stored must be recomputed each time regardless of whether they appear multiple times across the compressible Navier-Stokes. The pseudocode for the evaluation of the residuals is given in Figure \ref{fig:pseudocodeforRS}. This algorithm results in lengthy expressions for the residuals.

\begin{figure}[h]
\noindent\fbox{%
\begin{varwidth}{\dimexpr\linewidth-2\fboxsep-2\fboxrule\relax}
\begin{algorithmic}
\STATE {\texttt{Residual (mass) = $ \textrm{discretisation for  } \partial{\rho u_j}/\partial{x_{j}} + \rho \left(\textrm{arrays of } \partial{u_j}/\partial{x_{j}} \right) + \textrm{discretisation for } u_j \partial{ \rho}/\partial{x_{j}}$}}
\STATE{\texttt{Residual (momentum) = ...}}
\STATE{\texttt{Residual (Energy) = ...}}
\end{algorithmic}
\end{varwidth}
}
\caption{Pseudo-code for the evaluation of residuals in the RS algorithm.}
\label{fig:pseudocodeforRS}
\end{figure}

\subsubsection{Store Some (SS)}
In the SS algorithm, the derivatives (except for the velocity derivatives $\partial u_i/\partial x_j$) are evaluated on-the-fly and stored as thread-/process-local variables. These variables can then be re-used across different equations. The residuals are computed using a combination ofthese local variables and the stored arrays for the velocity gradient evaluation.
This algorithm requires storing 54 thread-/process-
local variables, as opposed to 54 global arrays in BL. The computational intensity of this algorithm is lower than RS algorithm. No attempt was made to sort the evaluation of local variables (e.g. grouping all the derivatives based on
coordinate direction or grouping the derivatives according to velocity components).

\subsection{Minimal storage algorithms}
We now consider variants of the BL algorithm that do not store any partial derivatives as arrays. They are either
recomputed on-the-fly, or evaluated and stored as thread-/process-local variables. No extra arrays
are used in these algorithms and three different versions are considered.

\subsubsection{Recompute All (RA)}
This algorithm follows a similar approach to that of the RS algorithm. The difference is that
the first derivatives of velocities and their dependant mixed-derivatives are also recomputed
on-the-fly. This results in one single large  expression to evaluate each residual and all the residuals are
evaluated in a single subroutine.

\subsubsection{Store None (SN)}
All the partial derivatives are evaluated on-the-fly and stored to 63 thread-/process-local variables, which
can be re-used across the equations. Similar to SS algorithm, the evaluation of derivatives
are not sorted. The residuals are computed using these local variables.

\subsubsection{Store None version 2 (SN2)}
This follows the same approach as the SN algorithm. However, the derivatives are grouped based on velocity
components such that all the derivatives using $u_i$ are grouped based on the index $i$, and the remaining derivatives are placed in another group.
First, the groups that contain velocity derivatives are evaluated in the order of the index $i$, and then the non-velocity group is evaluated.
The residuals are computed in a
similar fashion to that of the SN algorithm. This algorithm not reported in \cite{Jammy_etal_InPress} is used to investigate the effects of data
locality on different architectures, as a GPU's performance is known to be affected by data locality \cite{Ashby_etal_2010}.

\subsection{Source code generation for various architectures}
Newer hardware has the potential to reduce the runtime of existing numerical modelling software. However, most models are not in a position to readily exploit such architectures to their full potential. Porting often requires a non-trivial rewrite of the source code and newer architectures might also arrive during this process, resulting in further challenges for the numerical modeller. In an effort to future-proof models in the face of evolving hardware, a new finite difference modelling framework OpenSBLI, was introduced by \cite{Jacobs_etal_2017}. OpenSBLI automatically generates C code (which performs the discretisation, computes the solution and diagnostic fields, etc) from a high-level problem specification, using basic components of the symbolic Python package
SymPy \cite{sympy}. As the numerical discretisation is performed symbolically by the OpenSBLI framework, generating
the source codes for the different algorithms requires high-level information, such as the derivatives to store (e.g. the velocity
components for the SS and RS algorithms), and whether to evaluate derivatives to thread-/process-local variables on-the-fly or recompute them. The ease at which the algorithms can
be modified is one of the advantages of the automatic source code generation framework.

To target different back end architectures, we use the source-to-source translation capabilities of the OPS library
\citep{Mudalige_etal_2014, Reguly_etal_2014, Giles_etal_2015} to tailor the generated code towards different
parallel hardware backends, such as MPI for CPUs, CUDA and OpenCL for GPUs, and OpenACC for heterogeneous hardware
configurations. OpenSBLI writes out C code that is compliant with the OPS API. The backend models
used in the current work are described in section \ref{sect:hardware}.

A summary of the algorithms is presented in Table \ref{table:memory_usage} with the operation count per
timestep. This is obtained by counting the number of operations
for each expression in the algorithm using the \texttt{count\_ops} functionality in SymPy and adding them together. This sum is then
multiplied by the total number of grid points ($256^{3}$ used in this work) and the number of sub-stages in a Runge-Kutta iteration (3 in the current work)
to arrive at the operation count per timestep.

\begin{table}[!h]
   \begin{center}

      \begin{tabular}{|c|c|c|c|c|c|c|}
      \hline
        & BL & RS & SS & RA & SN &SN2 \\
      \hline
      Local variables & 0 & 0 & 53 & 0 & 63 & 63\\
      Extra arrays & 63 & 9 & 9 & 0 & 0 &0\\
      Operations ($\times 10 ^ {9}$) &48 & 68 & 54& 145& 69& 69 \\
      \hline
      \end{tabular}
      \caption{Number of process-/thread-local variables, work arrays and operation count used by each algorithm, for a three stage Runge-kutta time step.}
      \label{table:memory_usage}
   \end{center}
\end{table}
Operation count is an indication of the compute intensity of the algorithms. It should also be noted that,
as different compilers optimise differently, the actual operation count performed by the resulting binary executable would vary.
Expensive divisions are avoided by evaluating rational numbers and the inverse of constants at the start of the program. A complete list of the optimisations
performed is reported in \cite{Jammy_etal_InPress}.

\section{Hardware and execution}\label{sect:hardware}
OpenSBLI uses OPS's source-to-source translation capabilities to tailor and optimise the algorithm for their execution on various architectures. On the CPUs and KNLs a variety of programming models such as MPI, MPI+OpenMP, OpenACC, inlined MPI, and Tiled may be used.
As it is beyond the scope of this paper to compare different programming models,
we select the MPI model on the CPUs and KNL. This yields a typical manually-parallelised solver with
a similar performance for parallelisation as reported in \citep{Mudalige_etal_2014}.

Similarly, for GPUs one can use CUDA, OpenCL, OpenACC or MPI\_CUDA for multiple GPUs.
Our experience \cite{Jammy_etal_2015} is that CUDA performs as well as OpenCL, so the CUDA backend is used
for the execution of algorithms on GPUs.

The CPU simulations are performed on the UK National Supercomputing Service (ARCHER).
This is a Cray XC30 system with each CPU compute node comprising two Intel 12-core E5-2697 v2 Ivy Bridge CPUs,
each with a clock frequency of 2.7 GHz and 64 GB of RAM shared between them \citep{EPCC_2016}.

The KNL simulations were also performed on ARCHER since it comprises of Intel Xeon Phi Knights Landing (KNL) 7210 processors
(with a clock frequency of 1.30 GHz, 16 GB of high-bandwidth MCDRAM, and 96 GB of DDR4 RAM) with 64 cores.

In the case of the GPU runs, all simulations were performed
on a single NVIDIA Tesla K40c installed on a local desktop computer. This comprises 2,880 CUDA
stream cores, and runs at a clock speed of 745 MHz with 12 GB of on-board memory \citep{NVIDIA_2013}.

\subsection{Performance measurement}
For measuring the performance and energy consumption by the algorithms on various architectures,
the performance measurement libraries for those architectures were used.
On the CPU and KNL, the PAT MPI library (v 2.0.0) \citep{Bareford_2016}  was used. PAT MPI library
uses the Cray PAT Application Programming Interface (API), giving easy control and requiring minimal modifications
to the source code to collect the performance parameters. It should be noted that the actual
performance measurement is done by the Cray PAT library and PAT MPI library calls ensure that the number of MPI processes
reading the hardware performance counters is minimised, and the master (rank 0) collects and outputs the performance data \cite{Bareford_2016}.

Using the PAT MPI library involves adding calls to the \texttt{pat\_mpi\_monitor} in the
source code at the point where performance measurements should be taken and the quantities that are to be monitored are
controlled by a Makefile. More details can be found in \cite{Bareford_2016,Bareford_2015a, Bareford_etal_2016}.

In the present work, for all the algorithms on CPU and KNL, \texttt{pat\_mpi\_monitor} was added at the
start and end of each timestep. The simulation runtime,  instantaneous power (W) and  cumulative  energy (J) were recorded
at the \texttt{pat\_mpi\_monitor} function locations, as shown in Listing \ref{list2}.
The measurement resolution on the XC30 system was 0.1 seconds \citep{Bareford_2016} which was fine enough for the
time between iterations in the present simulations.

\begin{lstlisting}[caption = {Pseudocode showing the locations of energy monitoring calls}, captionpos=b, label= {list2}]

for (int iter =0; iter< niter; iter++){ // main time loop
    pat_mpi_monitor(1);
    for (int stage=0; stage< 3; stage++){
        /*Evaluate the residuals..
         */
    }
    pat_mpi_monitor(2);
}
\end{lstlisting}

On the GPU the NVIDIA Management Library (NVML) \citep{NVIDIA_2015} was used. The function \texttt{nvmlDeviceGetPowerUsage}
was inserted into the code at the
start and end of each iteration to measure the power consumption.

\subsection{Compilation}
After the insertion of performance measurement calls, the code was tailored to the target
hardware backends using OPS's source-source translation capabilities. For the CPU and KNL architectures the Intel C/C++ compiler (v 17.0) was used, and the NVIDIA C compiler (v 8.0.44) was used for the GPU.
On all architectures we used the -O3 compiler optimisation flag combined with architecture-specific flags
(-xHost, -xMIC-AVX512 and sm\_35 compute architecture for CPU, KNL and GPU, respectively) to generate binary code specific to
each architecture. These are summarised in Table \ref{tab:compilers}. All the dependant libraries and the algorithms are compiled using
the same optimisation flags on each architecture.

\begin{table}[h!]
\centering
\begin{tabular}{|l|l|l|l|}
\hline
& \multicolumn{1}{c|}{CPU} & \multicolumn{1}{c|}{KNL} & \multicolumn{1}{c|}{GPU} \\ \hline
Compiler & \multicolumn{2}{l|}{Intel C/C++ compiler v17.0} & NVIDIA C compiler v8.0.44 \\ \hline
\begin{tabular}[c]{@{}l@{}}Power /energy\\ measurement\end{tabular} & \multicolumn{2}{l|}{\begin{tabular}[c]{@{}l@{}}Cray PAT v 6.3.0\\ PAT\_MPI\_LIB v 2.0.0\end{tabular}}  & NVML \\ \hline
\multicolumn{1}{|c|}{\multirow{2}{*}{Optimisation flags}} & \multicolumn{3}{c|}{-O3} \\ \cline{2-4}
\multicolumn{1}{|c|}{} & -xHost & -xMIC-AVX512 & \begin{tabular}[c]{@{}l@{}}-gencode arch=compute\_35\\ ,code=sm\_35\end{tabular} \\ \hline
Backends & \multicolumn{2}{l|}{MPI} & CUDA \\ \hline
Peak Flops&  518 G Flops & 3+T Flops & 1.43 T Flops\\ \hline
\end{tabular}
\caption{Compilers and optimisation flags used for the different architectures.}
\label{tab:compilers}
\end{table}

\section{Results}\label{sect:results}
To evaluate the performance of the algorithms, a 3D compressible Taylor-Green vortex problem was simulated. This is a suitable and widely-used test case as it has simple initial and boundary conditions, but is complex enough to thoroughly validate the numerical method and its accuracy. It starts with a sinusoidal velocity field and develops into turbulent flow generating smaller and smaller vortices. The equations were solved in a 3D cube ($0 \leq x_0 \leq 2\pi L$, $0 \leq x_1 \leq 2\pi L$, and $0 \leq x_2 \leq 2\pi L$) with periodic boundary conditions in all directions \citep{DeBonis_2013,BullJameson_2014}. The initial conditions are given by

\begin{equation}
   u_0(x_0, x_1, x_2, t=0) = \sin\left(\frac{x_0}{L}\right)\cos\left(\frac{x_1}{L}\right)\cos\left(\frac{x_2}{L}\right),
\end{equation}
\begin{equation}
   u_1(x_0, x_1, x_2, t=0) = -\cos\left(\frac{x_0}{L}\right)\sin\left(\frac{x_1}{L}\right)\cos\left(\frac{x_2}{L}\right),
\end{equation}
\begin{equation}
   u_2(x_0, x_1, x_2, t=0) = 0,
\end{equation}
\begin{equation}
   p(x_0, x_1, x_2, t=0) = \frac{1}{\gamma \mathrm{M}^2} + \frac{1}{16}\left(\cos\left(\frac{2x_0}{L}\right)+\cos\left(\frac{2x_1}{L}\right)\right)\left(2 + \cos\left(\frac{2x_2}{L}\right)\right).
\end{equation}
In all the simulations, $\mathrm{Re}$ = 1,600, $\mathrm{Pr}$ = 0.71, $\mathrm{M}$ = 0.1, and $\gamma$ = 1.4. The reference quantities $L$, $u_{\mathrm{ref}}$ and $\rho_{\mathrm{ref}}$ were set to 1.0, and the reference temperature $T_{\mathrm{ref}}$ was evaluated using the equation of state (equation \ref{eq:Tinprimitive}).

The simulation was set-up using $256$ grid points in each direction. The BL, RS, SS, RA and SN algorithms
are validated by \cite{Jammy_etal_InPress, Jacobs_etal_2017} using a Cray compiler on the ARCHER CPU node.
The results of the algorithms using the current compilers on different architectures are compared to the solution dataset of
\citep{Jammy_etal_2016_TGV_enstrophy_ke_dataset}. The minimum and maximum error in the solution variables between
\citep{Jammy_etal_2016_TGV_enstrophy_ke_dataset} and the current runs was found to be of the order of machine precision on
all three architectures.

To evaluate the energy efficiency of the algorithms, the simulations are run for 500 iterations with energy measurement enabled.
Each algorithm is repeated five times on each architecture and the performance results presented here are
averaged over the five runs. In the following sections
the runtime, power and energy usage of the algorithms are investigated for each architecture.

\subsection{CPUs}
The CPU simulations are run using 24 MPI processes on a single ARCHER compute node. Table \ref{table:runtimes_MPI_CPU} show the runtimes and speed-up relative to the BL algorithm on CPU. The variation between the runtimes for different runs for each algorithm was less than 1\% from the
average.

\begin{table}[!ht]
\begin{center}
\begin{tabular}{|c|c|c|}
\hline
\textbf{Algorithm} & \textbf{Runtime (s)} & \textbf{Speed-up} \\
    \hline
BL & 1216.39 & 1.00\\
%     \hline
RS & 715.51 & 1.70\\
%     \hline
SS & 690.30 & 1.76\\
%     \hline
RA & 558.70 & 2.18\\
%     \hline
SN & 549.73 & 2.21\\
%     \hline
SN2 & 559.86 & 2.17\\
    \hline
\end{tabular}
\caption{Runtime (in seconds) and speed-up (relative to BL) for all algorithms on the CPU using 24 MPI processes on ARCHER.}
\label{table:runtimes_MPI_CPU}
\end{center}
\end{table}

All other algorithms outperform the BL case and a speed-up of ($\sim$1.7-2.2) relative to BL was achieved.
Similar to the findings of \citep{Jammy_etal_InPress}, it is worth noting that the SS algorithm achieved the lowest runtime on the same ARCHER CPU hardware. In the present case the runtimes of the minimal storage algorithms (SN, RA and SN2) are within 2\% of each other. The difference in the runtime relative to the results reported by \citep{Jammy_etal_InPress} was likely due to different compilers being used (the Cray C compiler version 2.4.2 in \citep{Jammy_etal_InPress}, versus the Intel C compiler in the present work), which caused different optimisations to be performed at compile-time.

\begin{figure}[!ht]
\begin{center}
\includegraphics[width=0.49\columnwidth]{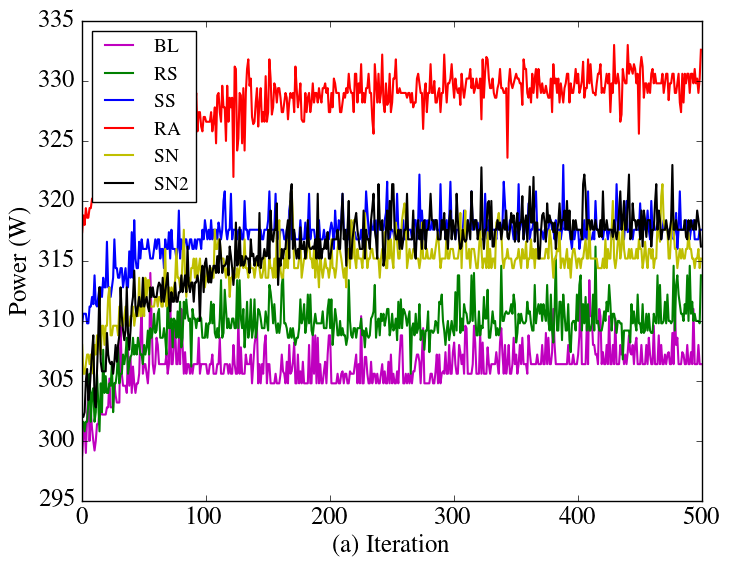}
\includegraphics[width=0.49\columnwidth]{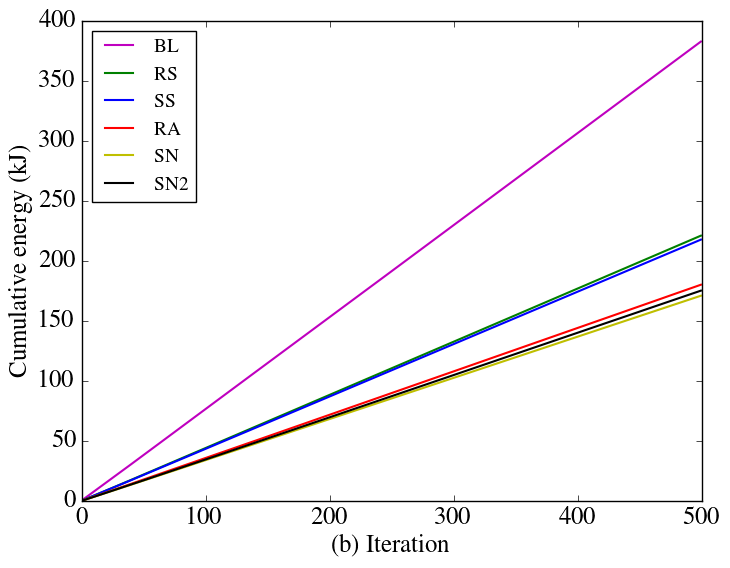}
\caption{(a) Power consumption per iteration (W), and (b) cumulative energy (kJ) for the different algorithms on the CPU node.}
\label{fig:power_cpu}
\end{center}
\end{figure}

Figure \ref{fig:power_cpu}(a) shows the power consumption of the algorithms for each iteration.
The steady rise in power at early times, which is common to all runs, is due to
Dynamic Voltage and Frequency Scaling (DVFS) seen in other studies of power consumption on CPUs \citep{Bareford_etal_2016}.
The BL algorithm consumes the least amount of power per iteration whereas RA consumes the most.
The higher power consumption for the compute intensive algorithms is expected due to the larger number of operations being performed. Note that SN and SN2 consume less power than RA since they reuse some of the thread-/process-local variables. With respect to the intermediate storage algorithms, RS consumes less power relative to SS. This is quite different to the behaviour of the
minimal storage algorithms (where more calculations lead to increased power consumption), and demonstrates that in order to be energy efficient, a balance needs to be achieved between the number of computations performed and the amount of local-storage used.

Figure \ref{fig:power_cpu}(b) shows the evolution of cumulative energy for each algorithm. As this represents
both power consumption and run-time, the lower the value, the more energy efficient the algorithm is.
In this figure the different classes of algorithms can be easily identified, irrespective of the power
consumption behaviour. BL uses the highest amount of energy, whereas all the compute intensive algorithms are close
to each other, consuming the least amount of energy. The intermediate storage algorithms consume more
energy than the compute-intensive algorithms but still far less energy than BL. The energy saving of an algorithm, defined as the ratio of energy used by the BL algorithm to the energy
used by that algorithm are 1.0, 1.73, 1.76, 2.12, 2.24, 2.18 for  BL, RS, SS, RA, SN and SN2 respectively.

From the performance of different algorithms on the CPU we can conclude that:
(a) the more computations that an algorithm performs with minimal read/write to RAM, the less energy is consumed;
(b) energy consumption of an algorithm is directly proportional to the speed-up relative to BL;
(c) the traditional way of writing large-scale finite difference codes (i.e. BL) is not optimal both in terms of runtime and energy consumption;
(d) high power consumption by an algorithm does not necessarily imply that the algorithm is energy inefficient; and
(e) improving the data locality of the thread-/process-local variables has negligible effect on both runtime
and energy consumption, which is expected as CPUs have relatively large caches.

\subsection{KNLs}
Simulations were run in parallel on the KNL using 64 MPI processes. Hyper-threading was not
enabled, in order to keep the configuration on the KNL similar to that of the CPU. It was also found to have
negligible effect on runtime in this case. As the memory used by the algorithms for the grid size considered can fit in the 16GB of on-chip high bandwidth memory (MCDRAM memory),
the entire MCDRAM is utilised in the cache memory.

\begin{table}[h!]
\begin{center}
\begin{tabular}{|c|c|c|}
\hline
\textbf{Algorithm} & \textbf{Runtime (s)} & \textbf{Speed-up} \\
    \hline
BL & 739.61 & 1.00\\
    \hline
RS & 425.02 & 1.74\\
    \hline
SS &  426.05 & 1.74\\
    \hline
RA & 415.59 & 1.78\\
    \hline
SN & 410.96 & 1.80\\
    \hline
SN2 & 401.99 & 1.84\\
    \hline
\end{tabular}
\caption{Runtime (in seconds) and speed-up (relative to BL) for all algorithms on the KNL using 64 MPI processes.}
\label{table:runtimes_MPI_KNL}
\end{center}
\end{table}

Table \ref{table:runtimes_MPI_KNL} gives the runtime of the algorithms for each run and the speed-up achieved
relative to BL. Similar to the CPU runs, the variation between individual runs of the same
algorithm is $\sim$ 1-2\% from the mean. A speed-up of $\sim$ 1.8 is achieved for all the algorithms relative to BL. No significant variation
in speed-up/runtime is seen between the low and intermediate storage algorithms; this might be due
to the relatively small grid size and also the use of high band-width memory.
However, the trend is similar to that seen in the CPU (i.e. BL is the slowest, the compute intensive algorithms are the
fastest, and intermediate storage algorithms are in-between). With increased grid size the differences might be more pronounced.
It is also interesting to note that the SN2 algorithm, which is the same as SN except that the local variable evaluation is reordered,
was 2.5\% faster than the SN algorithm, unlike the CPU case where SN2 was slower than SN.

\begin{figure}[!ht]
\begin{center}
\includegraphics[width=0.49\columnwidth]{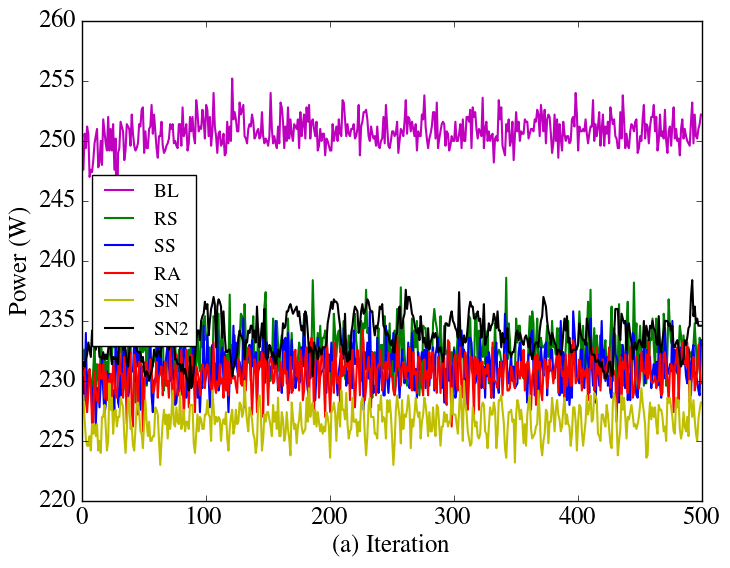}
\includegraphics[width=0.49\columnwidth]{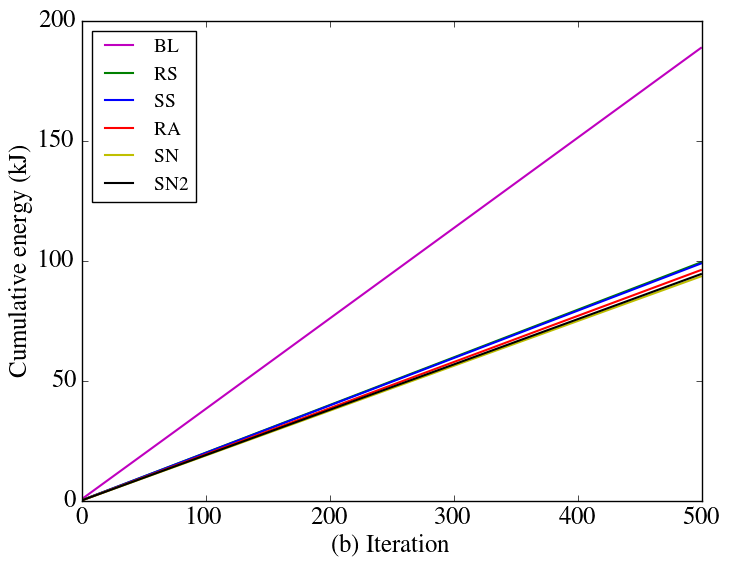}
\caption{(a) Power consumption per iteration (W), and (b) cumulative energy (kJ) for the different algorithms on the KNL.}
\label{fig:power_energy_knl}
\end{center}
\end{figure}

Figure \ref{fig:power_energy_knl}(a) shows the power consumption per iteration. Unlike CPUs, there is not much
effect of DVFS at early times. The BL algorithm consumes the highest amount of
power on KNL, which might arise from the exchanges between the MPI ranks, which can be reduced for hybrid
MPI and OpenMP codes. The other algorithms consume a similar amount of power per iteration.

Figure \ref{fig:power_energy_knl}(b) shows the evolution of cumulative energy consumption for each algorithm.
Similar to the CPU case, the BL algorithm consumes the highest amount of energy and all other algorithms consume $\sim$ 50\%
less. The energy savings are 1.0, 1.9, 1.91, 1.96, 2.02 and 2.0 for BL, RS, SS, RA, SN and SN2,
respectively.

The following can be concluded from the results:
    (a) BL is the worst performing in terms of energy and runtime;
    (b) the energy consumption is proportional to the speed-up achieved by the algorithm relative to BL;
    (c) the effects of re-ordering of derivatives in the SN2 are apparent and became more pronounced with larger grid sizes;
    (d) high power consumption by an algorithm does not necessarily imply that the algorithm is energy inefficient, and measuring power
    alone is not the best way to interpret the energy efficiency of an algorithm.

\subsection{GPUs}

\begin{table}[h!]
\begin{center}
\begin{tabular}{|c|c|c|}
\hline
\textbf{Algorithm}  & \textbf{Runtime (s)} & \textbf{Speed-up} \\
    \hline
BL & 496.29 & 1.00\\
    \hline
RS & 255.52 & 1.94\\
    \hline
SS & 231.25 & 2.15\\
    \hline
RA & 234.29 & 2.12\\
    \hline
SN & 297.68 & 1.67\\
    \hline
SN2 & 220.45 & 2.25\\
    \hline
\end{tabular}
\caption{Runtime (in seconds) and speed-up (relative to BL) for all algorithms on the NVIDIA Tesla K40 GPU using CUDA.}
\label{table:runtimes_CUDA}
\end{center}
\end{table}

Table \ref{table:runtimes_CUDA} shows the runtime of the algorithms on GPUs.
Similar to the CPU and KNL, all algorithms perform better than BL.
For the current grid size used, no specific trend can be found between low storage and intermediate
storage algorithms. The fact that RA is slower than SS, which is not seen on other
architectures, shows the difficulty of coding numerical methods on a GPU as these architectures
are sensitive to data locality. This can be inferred from the runtimes of SN and SN2,
with the only difference between them
being improving data locality which gives a $\sim$ 30\% reduction in runtime.
The order in which we write the evaluations while performing compute-intensive operations does indeed have
an effect on the runtime on a GPU as shown in \cite{Ashby_etal_2010} .

\begin{figure}[!ht]
\begin{center}
\includegraphics[width=0.49\columnwidth]{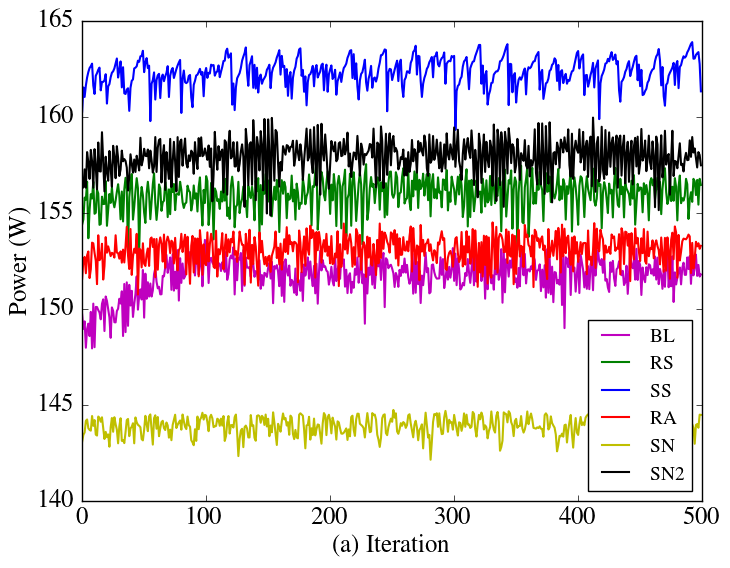}
\includegraphics[width=0.49\columnwidth]{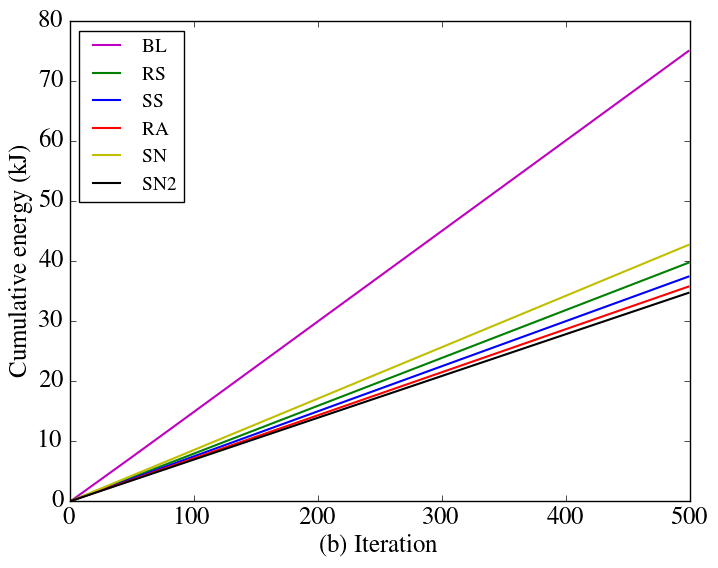}
\caption{(a) Power consumption per iteration (W), and (b) cumulative energy (kJ) for the different algorithms on the GPU.}
\label{fig:power_energy_cuda}
\end{center}
\end{figure}

Figure \ref{fig:power_energy_cuda}(a) shows the power consumed per iteration.
On the GPU, the SN algorithm uses less power due to the lack of data
locality. If we ignore the SN algorithm from the current discussion, then BL uses less power and interestingly SS
uses more power. The power usage by the intermediate storage algorithms is higher than that of the
most computationally-intensive algorithm (RA), again indicating that data locality has a critical impact on GPU performance.

Figure \ref{fig:power_energy_cuda}(b) shows the evolution of cumulative energy. Similar to the CPU and KNL,
the BL algorithm consumes the highest amount of energy. If we set aside the SN algorithm from the discussion, the 
SN2 algorithm uses the least amount of energy. Even though the runtime of RA is higher than the runtime of SS by 1.5\%,
the energy consumption of RA is less than SS by 5\%. This means that RA algorithm can be improved further by
writing the large expression in such a way that data locality is enhanced. The energy savings of the algorithms are
1.0, 1.89, 2.01, 2.1 and 2.16 for BL, RS, SS, RA and SN2 respectively, showing a similar trend
to the CPU and KNL.

We can conclude the following for GPU: (a) the performance of BL is inefficient both in terms of runtime and energy;
(b) data locality is inversely related to the power consumed;
(c) care should be taken while optimising algorithms on GPU to improve data locality;
(d) reading and writing to arrays should be minimised, even though the computational intensity of an algorithm increases;
(e) all the algorithms consume half the amount of energy relative to BL.

\section{Summary and conclusions}\label{sect:conclusion}
The route to exascale presents additional challenges to the numerical modeller. Not only is it
crucial to consider runtime performance of the algorithms that underpin finite-difference codes,
it is also important  for model developers to consider the energy efficiency of their codes.
This paper has highlighted  the benefits of using an automated code generation framework such as
OpenSBLI to readily vary the
computational and memory intensity of the finite difference algorithms in order to evaluate their
performance.

\begin{figure}[!h]
\begin{center}
\includegraphics[width=0.8\columnwidth]{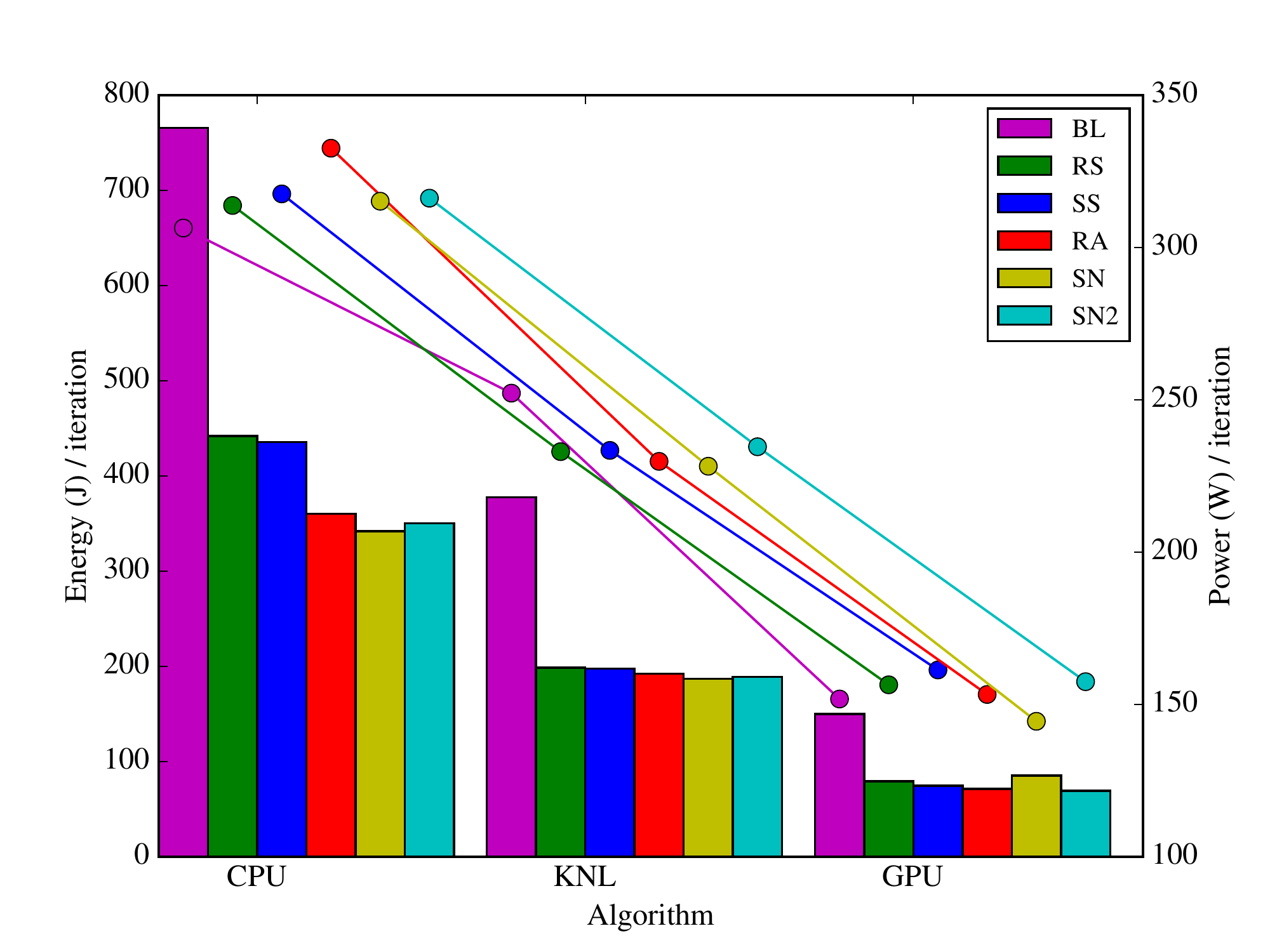}
\caption{Energy per iteration (bars) and the power consumed  per iteration (lines) on different
         architectures.}
\label{fig:energy_power_1}
\end{center}
\end{figure}

To summarise the findings, Figure \ref{fig:energy_power_1} shows the energy consumed per iteration and the
average power consumption per iteration for each algorithm  on the three architectures considered.
The CPU node use more power per iteration than a KNL or GPU.
GPUs use about 50\% less power than a CPU node and the KNL
systems are in-between. The trend is similar for energy consumption per iteration on three architectures, which
agrees with the findings of \cite{Hernandez2015}, in which a second order central finite difference scheme was
used for solving the acoustic diffusion model equation; CPUs use more energy and GPU use less and Xeon Phi (Knights Corner)
processors are in-between.

\begin{figure}[!ht]
\begin{center}
\includegraphics[width=0.8\columnwidth]{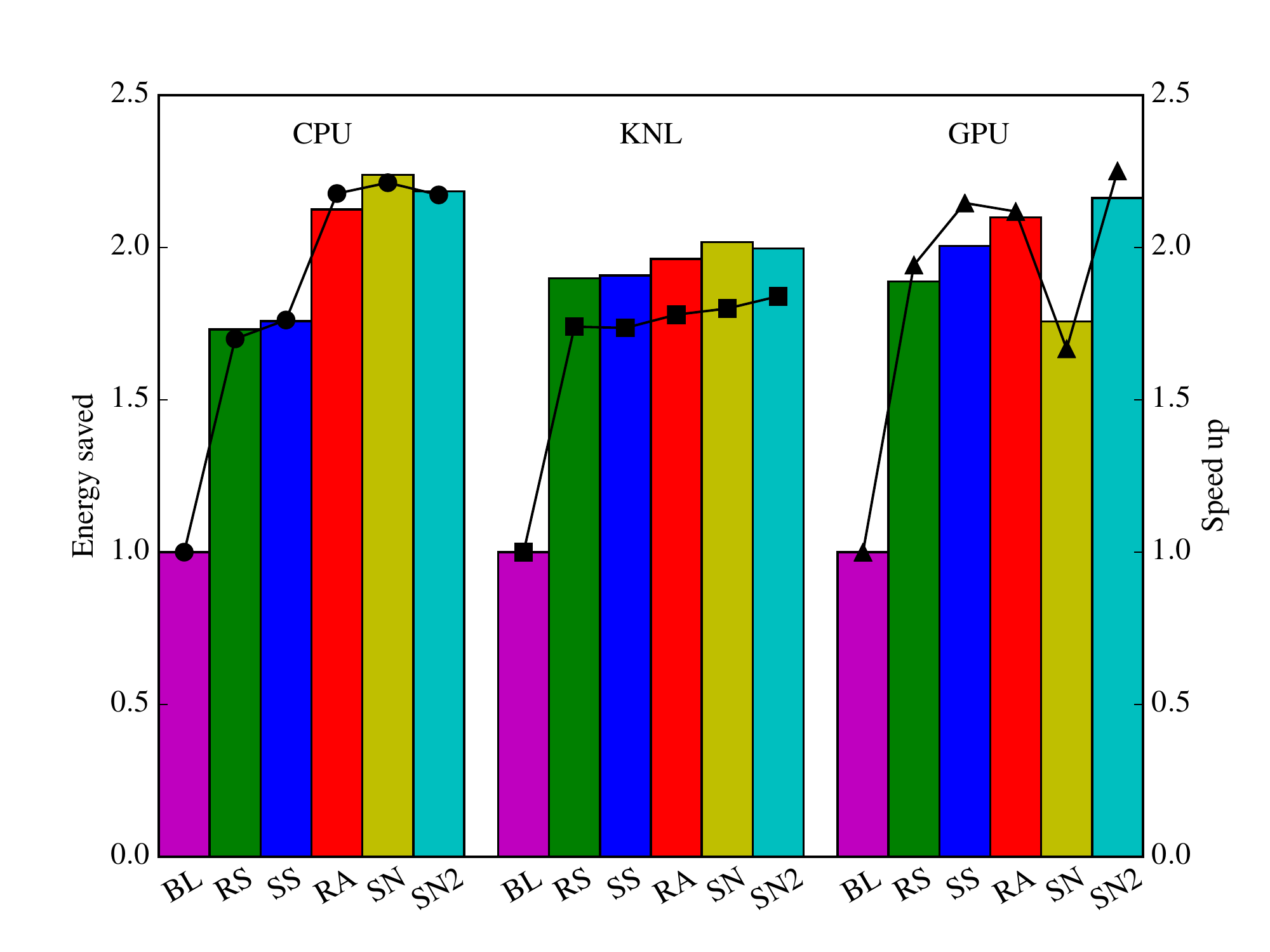}
\caption{Energy saved (bars) and the speed-up (lines) of algorithms relative to the BL algorithm for each
         architecture.}
\label{fig:energy_power_2}
\end{center}
\end{figure}

Figure \ref{fig:energy_power_2} shows a comparison of speed-up of the algorithms and the energy
saved by the algorithms on different architectures. The energy saved is in a direct
proportion with speed-up relative to the baseline algorithm.

The conclusions from the present work include:
(a) Across all architectures, the low storage/ high compute intensity algorithms are more energy efficient
and gives the best performance;
(b) Runtime reduction reduces the energy used by the algorithm;
(c) For all the algorithms considered CPUs are the least energy efficient and GPUs are the most energy efficient, with the KNL
architecture in-between;
(d) For the high compute intensity algorithms, the energy saving of KNL and GPU are $\sim$ 2 and $\sim$ 5
compared to the CPU node;
(e) For optimising the simulations on GPU, care should be taken to improve data locality.

\section{Acknowledgments}
SPJ and CTJ were supported by a European Commission Horizon 2020 project grant entitled ``ExaFLOW: Enabling Exascale Fluid Dynamics Simulations'' (grant reference 671571). DJL was supported by an EPSRC Centre for Doctoral Training grant (EPSRC grant EP/L015382/1). The data behind the results presented in this paper will be available from the University of Southampton's institutional repository. The authors acknowledge the use of the UK National Supercomputing Service (ARCHER), with computing time provided by the UK Turbulence Consortium (EPSRC grant EP/L000261/1). The NVIDIA Corporation kindly donated the Tesla K40c GPU used throughout this research.
Some of the CPU and KNL results were presented orally at the 29th Parallel Computational Fluid Dynamics (ParCFD 2017) conference.
\appendix


%% References

\section*{References}
\bibliographystyle{elsarticle-num}
\bibliography{energy-efficiency}






% \end{thebibliography}

\end{document}